\documentclass[useAMS,usenatbib]{mn2e}
\usepackage{graphicx}
\usepackage{amsmath}
\usepackage{fixltx2e}

\usepackage{amsmath}
\usepackage{amssymb}

\newcommand{\Msun}{\hbox{$\rm\thinspace \text{M}_{\odot}$}}

\newcommand{\ms}{$\,{\rm M}_\mathrm{\odot}$}

\newcommand{\diff}[2]{\frac{\partial (#1)}{\partial #2}}

\newcommand{\bi}[1]{\textbf{\textit{#1}}}

\title[Stellar populations including rotation]{Towards a unified model of stellar rotation II: Model-dependent characteristics of stellar populations}

\author[A. T. Potter, C. A. Tout, I. Brott]{Adrian~T.~Potter$^1$\thanks{E-mail: apotter@ast.cam.ac.uk}, Christopher~A.~Tout$^1$ and Ines~Brott$^2$\\
$^1$Institute of Astronomy, The Observatories, Madingley Road, Cambridge CB3 0HA\\
$^2$Institute for Astrophysics, Tuerkenschanzstr.17, AT-1180, Vienna, Austria}

\begin{document}
\date{Accepted 2012 March 17. Received 2012 March 9; in original form 2012 January 16}

\maketitle

\begin{abstract}

Rotation has a number of important effects on the evolution of stars. Apart from structural changes because of the centrifugal force, turbulent mixing and meridional circulation caused by rotation can dramatically affect a star's chemical evolution. This leads to changes in the surface temperature and luminosity as well as modifying its lifetime. Observationally rotation decreases the surface gravity, causes enhanced mass loss and leads to surface abundance anomalies of various chemical isotopes. The replication of these physical effects with simple stellar evolution models is very difficult and has resulted in the use of numerous different formulations to describe the physics. Using stellar evolution calculations based on several physical models we discuss the features of the resulting simulated stellar populations which can help to distinguish between the models.

\end{abstract}
\begin{keywords}
stars:evolution, stars:general, stars:rotation, stars:abundances, stars:chemically peculiar
\end{keywords}

\section{Introduction}

The effect of rotation on the internal physics of stars has been considered for many years \citep[e.g.][]{Kippenhahn70}. Rotation causes significant changes in the hydrostatic balance of the star \citep{Endal76}, thermal imbalance causes a meridional circulation current \citep{Sweet50} and differential rotation leads to shear instabilities \citep[e.g.][]{Spiegel70}. These all result in the mixing of angular momentum and chemical elements within the star leading to changes its surface properties such as the surface gravity, temperature, luminosity and chemical composition. Over the course of several decades, the physical formulations used to describe stellar rotation have proliferated \citep{Zahn92,Talon97,Meynet97,Heger00,Maeder05}. Whilst each new model has been suitably justified physically, there has been little observational data to back up claims of improved physical agreement. This leads to the possibility that any number of physical models can be chosen to produce a range of desired results which may or may not be accurate. This situation is worsened because the data required to constrain the models is scarce. However, with the observations of the VLT-FLAMES survey of massive stars \citep{Evans05,Evans06} and VLT-FLAMES Tarantula survey \citep{Evans10} it is now becoming possible to make such comparisons of different physical models and place some constraints on the formulations used. 

Comparing stellar models is still problematic because of the difficulty of isolating the effects of rotation from other physical and numerical differences in the results of other groups. \citet[][hereinafter referred to as Paper~1]{Potter11} presented {\sc rose}, a code capable of performing stellar evolution calculations with a number of different models of stellar rotation, eliminating any differences owing to other numerical or physical effects between different codes. 

In Paper~1 we considered the main differences between the evolution of individual stars under the assumptions of several popular models. In this paper we combine that analysis with the stellar population code, {\sc starmaker} \citep{Brott11}, to determine the difference in stellar populations that arise from two physical models. One is based upon \citet{Heger00} and has solely diffusive transport of angular momentum. The other is based on \citet{Talon97} and \citet{Maeder03} and has both diffusive and advective transport of angular momentum. The two different models have very different diffusion coefficients and there are marked differences in the results for individual stars. It is possible to get better agreement between the models under different criteria by adjusting the associated unknown constants but this leads to poorer agreement elsewhere.

One particular consequence of different input physics that we found in Paper~1 is that the mass dependence of the mixing is very different in each case. The models agree in that the total enrichment in low-mass stars ($M < 20$\ms) is much less than in high-mass stars ($M > 20$\ms). However, the enrichment found with each model is very different for low-mass stars despite reasonable agreement for high-mass stars. In Paper~1 we also concluded that the difference between the two models varies for different metallicities. For $Z=0.001$, the model based on \citet{Heger00} actually produces significantly more nitrogen enrichment in high-mass stars, particularly for slow and moderate rotators. We shall explore all of these features further in this paper.

In section~2 we briefly outline the details of the two codes, {\sc rose} and {\sc starmaker}. For full descriptions we refer the reader to \citet{Potter11} and \citet{Brott11}. We also describe the models under comparison. In section~3 we compare the stellar population predictions and consider the similarities and differences between the models. In section~4 we present our summary and conclusions.

\section{Input physics}
\label{code}

Let us present the physical ingredients of the numerical evolution code, {\sc rose}, and the population synthesis code, {\sc starmaker}, as well as the two models under comparison.

\subsection{RoSE}
\label{rose}

{\sc rose} is based on the Cambridge stellar evolution code, {\sc stars}, the first version of which  was  written by \citet{Eggleton71}. It has been modified and had its physics updated many times since. For details of the last major update see \citet{Eldridge09}. The code solves the four structure equations, seven chemical equations and now the angular velocity equation in single, implicit, Newton-Raphson iterative steps. We calculate $^1$H, $^3$He, $^4$He, $^{12}$C, $^{14}$N, $^{16}$O and $^{20}$Ne implicitly and 39 other isotopic abundances can be calculated explicitly. The structure equations including rotation and the implementation of the angular momentum evolution are described in Paper~1. We summarize the key components here. The equations for hydrostatic and thermal equilibrium are modified as by \citet{Endal76} and \citet{Meynet97}. A surface of constant pressure, $P$ is defined as $S_P$, while $V_P$ is the volume contained within $S_P$ and $r_P$ is the radius of a sphere with volume $V_P=4\pi r_P^3/3$. Mass conservation gives

\begin{equation}
\frac{d\,m_P}{d\,r_P}=4\pi r_P^2\rho,
\end{equation}

\noindent where $m_P$ is the mass enclosed within $S_P$ and $\rho$ is the density on the isobar which is assumed to be uniform on $S_P$ even when the star rotates differentially. Strong horizontal turbulence keeps relevant physical quantities uniform along isobars. The local gravity vector is

\begin{equation}
\label{geff}
\bi{g}_{\rm eff}=\left(-\frac{Gm_P}{r_P^2}+\Omega^2r_P\sin^2\theta\right) \bi{e}_{r}+\left(\Omega^2r_P\sin\theta\cos\theta\right)\bi{e}_{\theta},
\end{equation}

\noindent where $\Omega$ is the local angular velocity. The average of a quantity $q$ over $S_P$ is defined as
\begin{equation}
<q>\equiv\frac{1}{S_P}\oint_{S_P}q d\sigma,
\end{equation}

\noindent where $d\sigma$ is a surface element of $S_P$. The equation for hydrostatic equilibrium is
\begin{equation}
\frac{dP}{dm_P}=-\frac{Gm_P}{4\pi r_P^4}f_P,
\end{equation}

\noindent where 
\begin{equation}
f_P=\frac{4\pi r_P^4}{Gm_PS_P}<g_{\rm eff}^{-1}>^{-1}
\end{equation}

\noindent and $g_{\rm eff}\equiv |\bi{g}_{\rm eff}|$. The thermal equilibrium equation is

\begin{equation}
\frac{d \ln T}{d \ln P}=\frac{3\kappa P L_P}{16 \pi acGm_PT^4}\frac{f_T}{f_P},
\end{equation}

\noindent where $L_P$ is the total energy flux through $S_P$, $P$ is the pressure, $T$ is the temperature, $\kappa$ is the opacity, $a$ is the radiation constant, $c$ is the speed of light, $G$ is the gravitational constant and

\begin{equation}
f_T\equiv\left(\frac{4\pi r_P^2}{S_P}\right)\left(<g_{\rm eff}><g_{\rm eff}^{-1}>\right)^{-1}.
\end{equation}

\noindent Again, the non-rotating equation for stellar evolution has been preserved except for the multiplication by $f_T/f_P$. Of the two factors, $f_P$ deviates further from unity for a given rotation than $f_T$. Additional secondary effects of the reduced gravity must be taken into account when calculating quantities such as the pressure scale height and Brunt--V\"ais\"al\"a frequency. For the remainder of this paper we drop the subscript $P$ on the variables.

We use the mass-loss rates of \citet{Vink01} for massive stars although several other mass-loss rates are often preferred. The enhanced mass-loss rate resulting from rotation is given by

\begin{equation}
\dot{M}=\dot{M}_{\Omega=0}\left(\frac{GM}{r^2<g_{\rm eff}>}\right)^{\xi/2},
\end{equation}

\noindent where we take $\xi=0.45$. This is the same as the one used by \citet{Langer98} except that we have chosen the critical rotation rate such that $\Omega_{\rm crit}$ corresponds to $<g_{\rm eff}>=0$. For a more complete description of the critical rotation rate of stars we refer the reader to \citet{Maeder00} and \citet{Georgy11}.

The thermal flux, $F\propto g_{\rm eff}(\theta)$ \citep{VonZeipel24} strongly depends on co-latitude. This produces a thermal imbalance that drives a meridional circulation. We use a formulation based on energy conservation along isobars similar to \citet{Maeder98}. In spherical polar coordinates the circulation takes the form

\begin{equation}
\bi{U}=U(r)P_2(\cos\theta)\bi{e}_r+V(r)\frac{dP_2(\cos\theta)}{d\theta}\bi{e}_{\theta},
\end{equation}

\noindent where $U$ and $V$ are linked by continuity so that
\begin{equation}
V=\frac{1}{6\rho r}\frac{d}{dr}(\rho r^2U)
\end{equation}

\noindent and $P_2(x)=\frac{1}{2}(3x^2-1)$ is the second Legendre polynomial, and

\begin{equation}
\begin{split}
U=C_0\frac{L}{m_{\rm eff}g_{\rm eff}}\frac{P}{C_P\rho T}\frac{1}{\nabla_{\rm{ad}}-\nabla+\nabla_{\mu}}\\\left(1-\frac{\epsilon}{\epsilon_m}-\frac{\Omega^2}{2\pi G\rho}\right)\left(\frac{4\Omega^2r^3}{3Gm}\right),
\end{split}
\end{equation}

\noindent where $m_{\rm eff}=m\left(1-\frac{\Omega^2}{2\pi G\rho}\right)$, $\epsilon=E_{\rm nuc}+E_{\rm grav}$, the total local energy emission, $\epsilon_m=L/m$, $C_P$ is the specific heat capacity at constant pressure, $\nabla$ is the radiative temperature gradient, $\nabla_{\rm{ad}}$ is the adiabatic temperature gradient, $\nabla_{\mu}$ is the mean molecular weight gradient and $C_0$ is a calibration constant.

Differential rotation arises in stars because of hydrostatic structural evolution, mass loss and meridional circulation. This leads to hydrodynamic instabilities that redistribute angular momentum. The resulting turbulence is much stronger horizontally than vertically and so variables are assumed to be roughly constant over isobars. Specifically we can describe the angular velocity distribution by $\Omega=\Omega(r)$.

Taking into account all of the processes described in section~\ref{rose} we use, as the evolution equation for the angular velocity \citep{Zahn92},

\begin{equation}
\begin{split}
\label{main1}
\diff{r^2\Omega}{t}=\frac{1}{5r^2}\diff{r^4\Omega U}{r}+\frac{1}{r^2}\frac{\partial}{\partial r}\left({D_{\rm{shear}} r^4\frac{\partial\Omega}{\partial r}}\right)\\+\frac{1}{r^2}\frac{\partial}{\partial r}\left({D_{\rm{conv}} r^4\frac{\partial \Omega}{\partial r}}\right)
\end{split}
\end{equation}

\noindent and for the chemical evolution

\begin{equation}
\label{main2}
\frac{\partial c_i}{\partial t}=\frac{1}{r^2}\frac{\partial}{\partial r}\left(\left(D_{\rm{shear}}+D_{\rm{eff}}+D_{\Omega=0}\right)r^2\frac{\partial c_i}{\partial r}\right),
\end{equation}

\noindent where $c_i$ is the abundance of element $i$. The diffusion coefficient $D_{\rm conv}$ is non-zero only in convective zones and $D_{\rm shear}$ and $D_{\rm eff}$ are non-zero only in radiative zones. The coefficient $D_{\rm eff}$ describes the effective diffusion of chemical elements because of the interaction between horizontal diffusion and meridional circulation. The variables $U$, $D_{\rm shear}$ and $D_{\rm eff}$ differ between the two test cases as described in section~\ref{testmodels}.

\subsection{Starmaker}

{\sc starmaker} is a population synthesis code described by \citet{Brott11}. It was originally designed to work with the evolutionary models of \citet{Brott11b}. We have adapted it for use with {\sc rose} stellar evolution models. Based on a grid of evolutionary models, {\sc starmaker} interpolates for stellar properties given an initial mass, initial surface velocity and age. These are chosen at random according to user-defined distribution functions. Each simulated star is assigned a random orientation in space. The newly generated sample can subsequently be filtered according to observational selection effects to enable comparison with observed samples. Such effects are not applied in this paper because the differences we describe are not strongly affected when they are.

In this study we limit ourselves to models on the main sequence, so if an age beyond the main-sequence is assigned to the model it is excluded from the simulation. Our initial masses are distributed with a Salpeter initial mass function. The initial surface rotation velocity distribution is that of \citet{Dufton06} for Galactic B-type stars, a Gaussian function truncated at zero with mean $\mu = 175 {\rm km\,s^{-1}}$ and standard deviation $\sigma=94{\rm km\,s^{-1}}$.

\subsection{Test cases}
\label{testmodels}

\begin{figure}
\begin{center}
\includegraphics[width=0.48\textwidth]{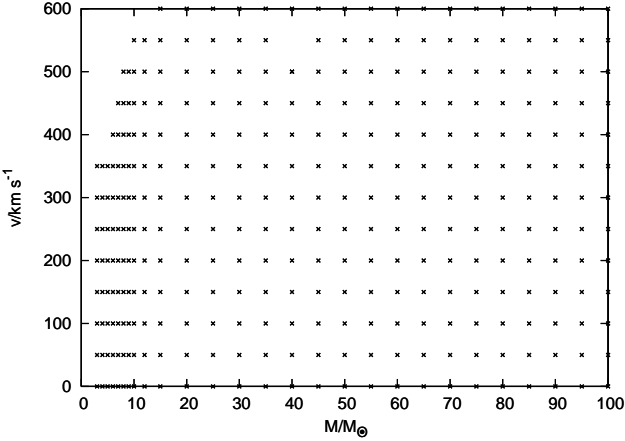}
\end{center}
\caption{Grid of initial models, in initial mass--initial equatorial velocity space, used for simulating stellar populations.}
\label{grid}
\end{figure}

We consider two models for comparison (cases~1 and~2 of Paper~I). For both models we evolve a grid of stars with masses between $3$ and $100$\ms\ and initial equatorial surface rotation velocities between $0$ and $600\,{\rm km\,s^{-1}}$. The zero age main sequence is the point of minimum luminosity at the onset of hydrogen burning. The masses computed are

\begin{align}
m/{\rm M}_\mathrm{\odot}\,\in\,&\{3,4,5,6,7,8,9,10,12,15,20,25,\nonumber\\ & 30,35,40,45,50,55,60,65,70,75,80,\nonumber\\ & 85,90,95,100\}
\end{align}

\noindent and for each mass the initial surface velocities used are

\begin{align}
v_{\rm ini}/{\rm km\,s^{-1}}\in\,&\{0,50,100,150,200,250,300,\nonumber\\ & 350,400,450,500,550,600\},
\end{align}

\noindent except when the rotation velocity would be too close to critical rotation to achieve numerical convergence. This becomes more difficult for stars less massive than $10$\Msun. Convergence can be achieved for a $10$\Msun\ star rotating faster than $95$\% of critical rotation, although the assumptions of the model are likely to become invalid this close to critical. For a $3$\Msun\ star the limit for convergence is close to $70$\% of critical rotation. Both the case~1 and case~2 models for each mass and initial surface velocity must reach the end of the main sequence for either of them to be used in the grid. The end of the main sequence is the point of maximum temperature before a star moves onto the Hertzsprung gap. Each model evolved is plotted in Fig.~\ref{grid}. For both models, the diffusion of angular momentum in convective zones is determined by the characteristic eddy viscosity given by mixing length theory such that $D_{\rm conv}=D_{\rm mlt}=\frac{1}{3}v_{\rm mlt}l_{\rm mlt}$. The position of the convective boundary is determined by the Schwarzschild criterion and although the code includes a model for convective overshooting, we do not use it in this paper. Unlike for Paper~1, we only consider the case in which the convective core tends to a state of solid body rotation. In section \ref{calibration} we examine the effects of changing the free parameters associated with the model. Models generated with this calibration are referred to as case~2$_{\rm b}$. In this paper we generate models with two different metallicities, Galactic and Large Magellanic Cloud (LMC), as defined by \citet{Brott11b}. Other than the initial composition, the input physics is the same for both metallicities. However, for clarity, we distinguish models that use LMC metallicity by referring to them with a superscript `Z' (e.g. case~${\rm 1^Z}$).

\subsubsection{case~1}

Our case~1 model uses the formulation for $D_{\rm shear}$ of \citet{Talon97},
\begin{equation}
D_{\rm shear}=C_0\frac{2 Ri_{\rm c}\left(r\frac{d\Omega}{dr}\right)^2}{N_T^2/(K+D_{\rm h})+N_{\mu}^2/D_{\rm h}}.
\end{equation}

\noindent where

\begin{equation}
N_T^2=-\frac{g_{\rm eff}}{H_P}\left(\frac{\partial\ln\rho}{\partial \ln T}\right)_{\!\!P,\mu}\left(\nabla_{\rm ad}-\nabla\right)
\end{equation}
\noindent and
\begin{equation}
N_{\mu}^2=\frac{g_{\rm eff}}{H_P}\left(\frac{\partial\ln\rho}{\partial\ln\mu}\right)_{\!\!P,T}\frac{d \ln\mu}{d \ln P}.
\end{equation}

\noindent As for paper~1 we follow \citet{Maeder03} by taking the critical Richardson number, $Ri_{\rm c}=(0.8836)^2/2$. We have also chosen $C_0$ so that we reproduce the terminal-age main-sequence (TAMS) nitrogen enrichment of a $40$\ms$\,$ star initially rotating at $270{\,\rm km\,s^{-1}}$ with Galactic composition given by \citet{Brott11b}. The effective diffusion coefficient $D_{\rm eff}$ is

\begin{equation}
D_{\rm eff}=\frac{|rU|^2}{30 D_{\rm h}},
\end{equation}

\noindent and we take
\begin{equation}
D_{\rm h}=0.134 r\left(r\Omega V\left[2V-\alpha U\right]\right)^{\frac{1}{3}},
\end{equation} 

\noindent where
\begin{equation}
\alpha=\frac{1}{2}\frac{d(r^2\Omega)}{dr}.
\end{equation}

\subsubsection{case~2}

Our case~2 model is that of \citet{Heger00}. In this case $U=0$ because circulation is treated as a purely diffusive process. The details of the various diffusion coefficients are extensive so we refer the reader to the original paper. With their notation the diffusion coefficients are
\begin{equation}
D_{\rm shear}=D_{\rm sem}+D_{\rm DSI}+D_{\rm SHI}+D_{\rm SSI}+D_{\rm ES}+D_{\rm GSF}
\end{equation}

\noindent and
\begin{equation}
D_{\rm eff}=(f_{\rm c}-1)(D_{\rm DSI}+D_{\rm SHI}+D_{\rm SSI}+D_{\rm ES}+D_{\rm GSF}),
\end{equation}

\noindent where each $D_i$ corresponds to a different hydrodynamical instability. \citet{Heger00} take $f_{\rm c}=1/30$ and we use this too. We also use $f_\mu=0$. The consequences of this are discussed in Paper~1. Unless otherwise stated, we calibrate this model by scaling $D_{\rm ES}$, the dominant diffusion coefficient, so that the nitrogen enrichment of a 20\ms, solar metallicity star with initial surface angular velocity of $300\,{\rm km\,s^{-1}}$ is the same as for case~1 at the terminal-age main sequence.

\subsection{Stellar populations}

\begin{table}
\begin{tabular}{cccc}
\hline
Age/yr&Number of&Maximum mass&Maximum mass\\
 &excluded stars&(case~1)/${\rm M}_\mathrm{\odot}$&(case~2)/${\rm M}_\mathrm{\odot}$\\
\hline
$5\times 10^6$&$250625$ & $42.2$ & $41.6$\\
$10^7$&$831418$&$21.6$ & $22.3$\\
$2\times 10^7$&$1681936$ & $14.1$ & $13.7$\\
$5\times 10^7$&$3550414$ & $8.2$ & $7.4$\\
\hline
\end{tabular}
\caption{
\label{pops}
The properties of different single-aged stellar populations used in section \ref{results}. The original size of the population in each case is $10^7$ stars. Each population is generated by an instantaneous burst of star formation at $t=0$. The first column shows the age of the simulated population. The second column shows the number of stars that have reached the end of the main sequence and so are excluded from the sample. The third and fourth columns show the mass of the most massive star remaining in the sample at the given age for case~1 and case~2 respectively.}
\end{table}

Throughout this paper we use a variety of populations at different ages with different star formation histories. The main reason for this is that, as a population ages, the mass of the most massive stars remaining in the main-sequence population decreases, whilst stars much less massive than the maximum mass have not had sufficient time to produce significant nitrogen enrichment. The combination of these tendencies allows us to follow how the amount of enrichment varies with mass. This applies specifically to clusters in which we expect the range of ages of the stars to be small compared to the age of the cluster. It is important to note that rotation can significantly affect the upper bound to the mass of stars in the population. The populations used in this paper are listed in table (\ref{pops}). For the figures in section \ref{results} we compare the data using 2D histograms for which we have separated the data into a grid of $50 \times 50$ bins. The number of stars in each bin divided by the total number of stars in case~1 is $n_1$ and similarly $n_2$ for case~2. In each comparison we reduce the size of the larger population to be the same size as the other by randomly removing stars.

\section{Results}
\label{results}

\begin{figure*}
\begin{center}
\hspace*{-2.2cm}
\includegraphics[width=1.2\textwidth]{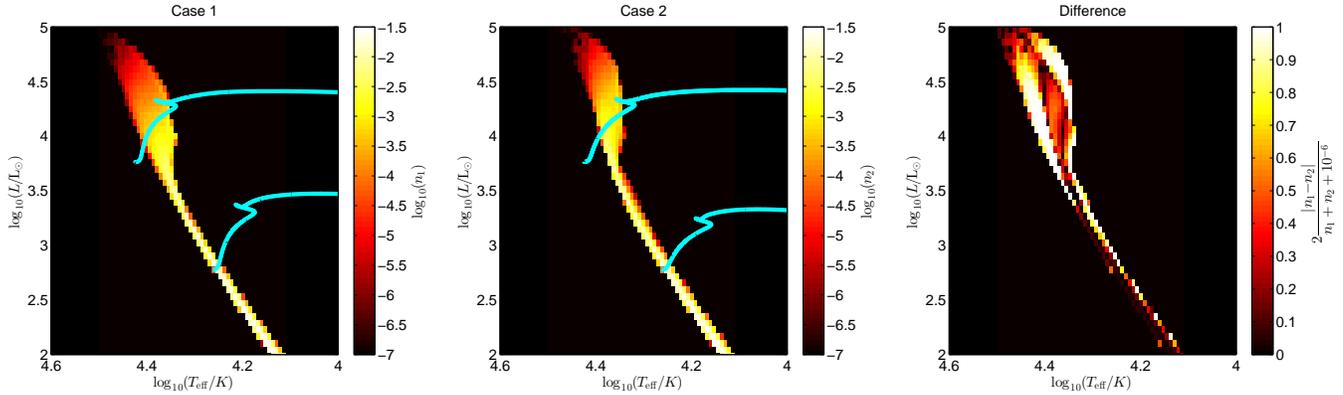}
\end{center}
\caption{Hertzsprung-Russell diagrams for a population of stars at age $2\times 10^7{\rm \,yr}$. In each case we have plotted the evolution of $5$\ms\ and $10$\ms\ stars initially rotating at $300\,{\rm km\,s^{-1}}$. There is some slight variation between the two cases at the end of the main sequence (highest luminosity) but the effect is small. Otherwise there is no obvious difference between the results produced in cases~1~and~2. When comparing the two populations, the addition of $10^{-6}$ in the denominator is to avoid division by zero in unpopulated bins.}
\label{2e7hr}
\end{figure*}


We simulated stellar populations at a number of ages and metallicities for each model and found a number of significant differences.

\subsection{The HR diagram}

When we look at the effect of the two models of rotation on stars in the HR diagram we find that there is very little difference between them. Whilst there is variation in the TAMS temperature and luminosities of the stars in each case, the difference is small and, for a single burst of star formation, only affects a handful of stars in the population at any given time. For most of a star's lifetime the predicted position in the HR diagram is sufficiently similar between the two cases that the difference in the population cannot be distinguished. Fig.~\ref{2e7hr} shows the HR diagrams for cases~1 and~2 for simulated clusters with an age of $2\times 10^7{\rm \,yr}$. Apart from slightly different degrees of broadening at the main-sequence turn off, there is no difference between the two cases. This is true at all ages and if we simulate a population of stars with continuous star formation we still find only slight distinctions between the two cases. This doesn't mean that mass determinations of rotating stars from their surface rotation, temperatures and luminosities are unaffected by the specific physics of the model. For individual models, the difference can be significant but the cumulative effect has little impact on the population as a whole. It is also important to note that the maximum mass of stars remaining in the sample varies between the cases because the main-sequence lifetimes are different. This means that care must be taken when identifying a cluster's age with respect to its most massive members if the cluster contains rapid rotators.

\subsection{Velocity distribution evolution}
\label{veldist}

\begin{figure*}
\begin{center}
\hspace*{-2.2cm}
\includegraphics[width=1.2\textwidth]{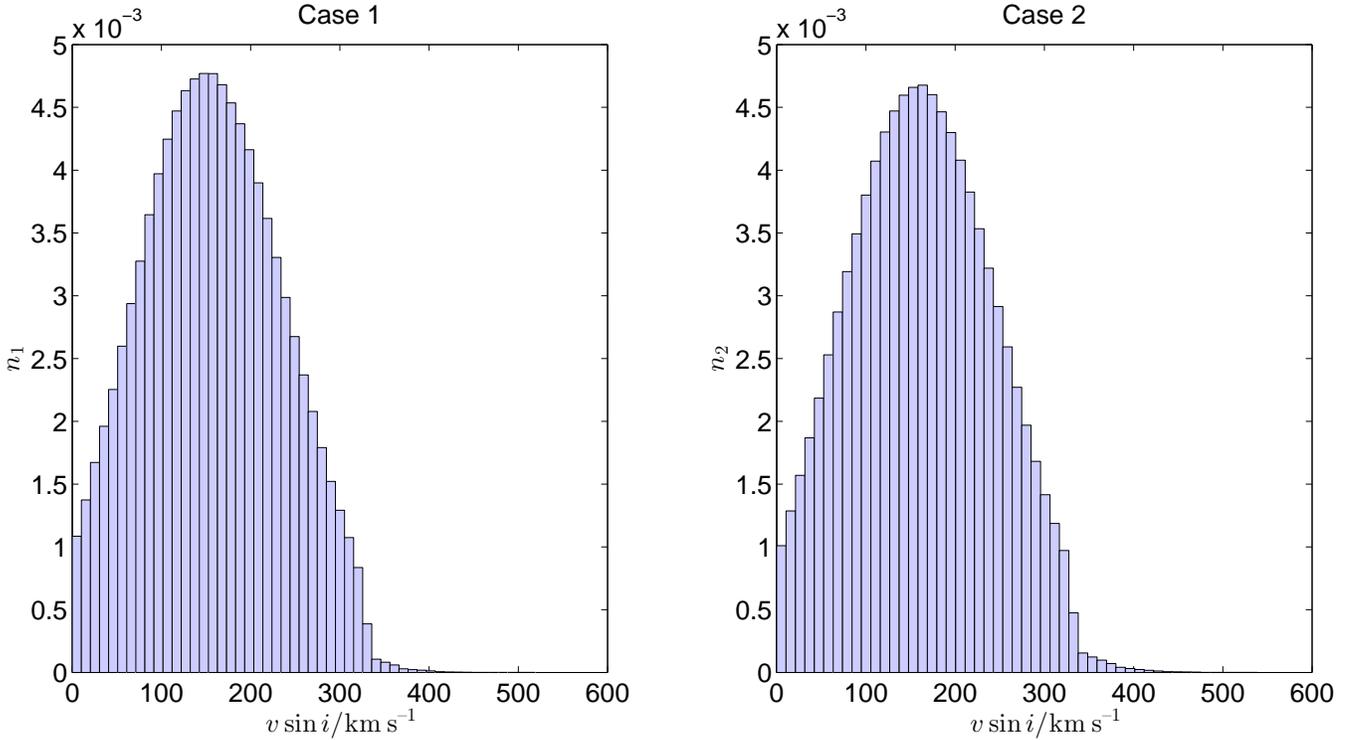}
\end{center}
\caption{Velocity distribution of stars in cases~1 and~2 both at an age of $2\times 10^7{\rm yr}$ for the populations described in table \ref{pops}. There is no marked difference between the two distributions.}
\label{2e7v}
\end{figure*}

Because of variations in the amount of mixing and the evolutionary timescale between the two cases, we might expect differences between the distribution of rotation rates as the populations evolve. In Fig.~\ref{2e7v} we plot the velocity distribution of the remaining stars in the single-aged populations at $2\times 10^7 {\rm yr}$. This is the typical shape of the distribution at all ages considered and we see that there is very little difference between the two cases. 

\subsection{The Hunter diagram}
\label{huntsec}

\begin{figure*}
\begin{center}
\hspace*{-2.2cm}\includegraphics[width=1.2\textwidth]{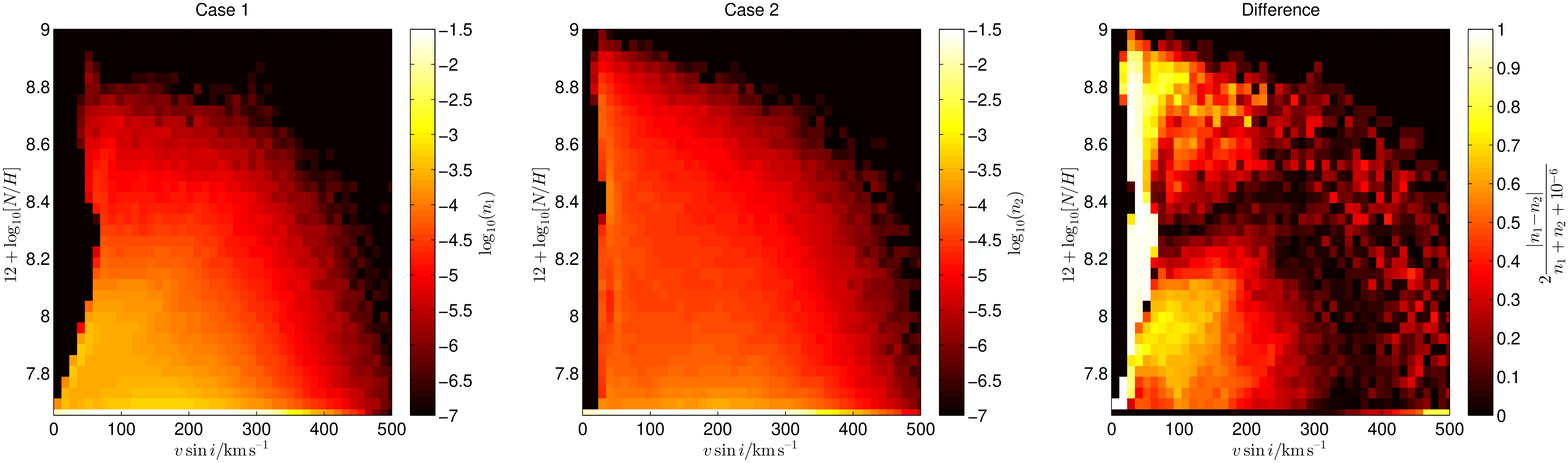}
\hspace*{-2.2cm}\includegraphics[width=1.2\textwidth]{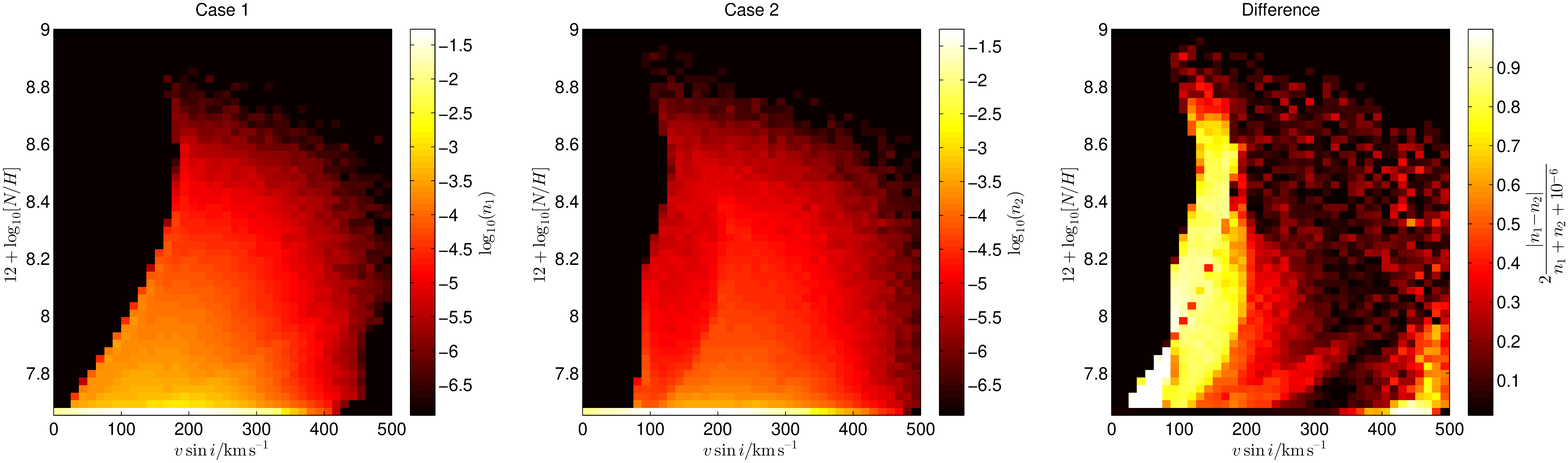}
\hspace*{-2.2cm}\includegraphics[width=1.2\textwidth]{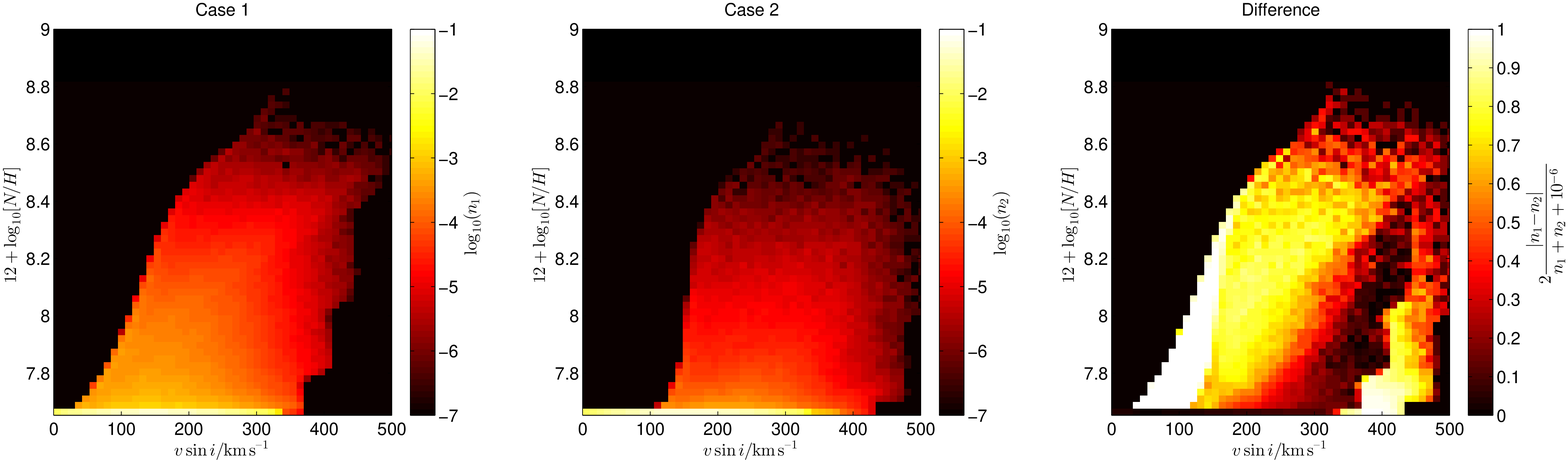}
\hspace*{-2.2cm}\includegraphics[width=1.2\textwidth]{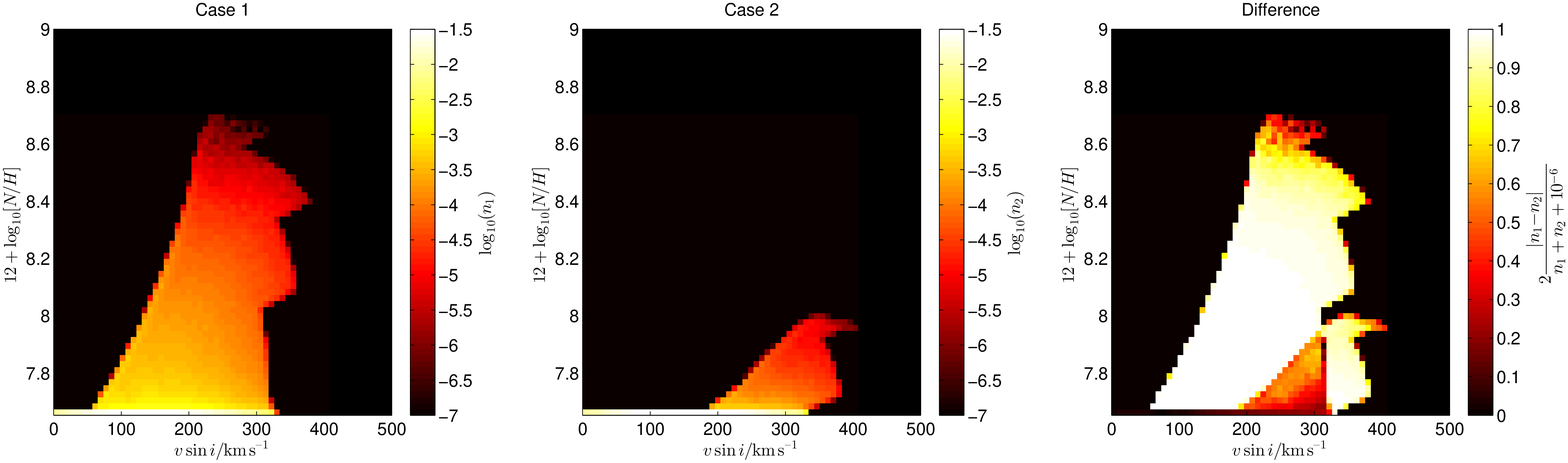}
\end{center}
\caption{Hunter diagrams for single-aged populations of stars. From top to bottom, the four rows of figures correspond to $5\times 10^6 {\rm yr}$, $10^7 {\rm yr}$, $2\times 10^7 {\rm yr}$ and $5\times 10^7 {\rm yr}$. At early times the two cases are similar with slightly more enrichment of the fastest rotators in case~2 and more enrichment of stars rotating more slowly than $200\,{\rm km\,s^{-1}}$ in case~1. By $2\times 10^7 {\rm yr}$ we see many more enriched stars in case~1 and by $5\times 10^7 {\rm yr}$ the amount of mixing in case~2 has dropped off dramatically. The jagged right hand edge of the populations is a result of the grid geometry and the mass-independence of the initial rotation velocity distribution. Neither affects the large difference we see in the populations at late times.}
\label{hunt}
\end{figure*}

The effect of rotation on the surface abundance of various isotopes is an extremely important tracer of the effects of rotation. In section \ref{veldist} we showed that, using alternate models of rotation, we find only small effects on the velocity distribution in stellar populations. We now consider the effect of rotation on the surface abundance of nitrogen. We could make similar conclusions about other chemical elements but their usefulness depends on the accuracy to which they can be measured and the availability of data. For example in Paper~1 we discussed the effect of rotation on the surface abundance of helium-3 but this is difficult to measure and so is not particularly useful in this discussion. \citet{Frischknecht10} and \citet{Brott11b} also consider how rotation is likely to affect the surface abundances of light elements. If we look at a plot of the surface abundance against surface rotation rate, commonly referred to as the Hunter diagram \citep{Hunter09}, for different ages (Fig.~\ref{hunt}) we see that there are some very clear differences between the two cases. At each age, the most massive stars remaining in the population dominate the enriched stars. Stars more massive than this have already evolved off the main sequence. The less massive stars in the population evolve more slowly and so have not had enough time to become enriched. At early times the populations are very similar except that case~1 predicts rather more enrichment for stars rotating slower than $200\,{\rm km\,s^{-1}}$ while case~2 predicts more enrichment of the most rapidly rotating stars. As the population ages, the amount of enrichment in case~1 stays roughly the same but the amount of enrichment in case~2 drops off slowly followed by a large drop between $2\times10^7 {\rm yr}$ and $5\times 10^7  {\rm yr}$. It is here, where only stars less massive than $8.2$\ms$\,$ remain, that the difference between the models is clearest. However, even at $2\times10^7 {\rm yr}$, we can see that there are far more enriched stars in case~1 than case~2 compared with earlier times.

\begin{figure*}
\begin{center}
\hspace*{-2.2cm}\includegraphics[width=1.2\textwidth]{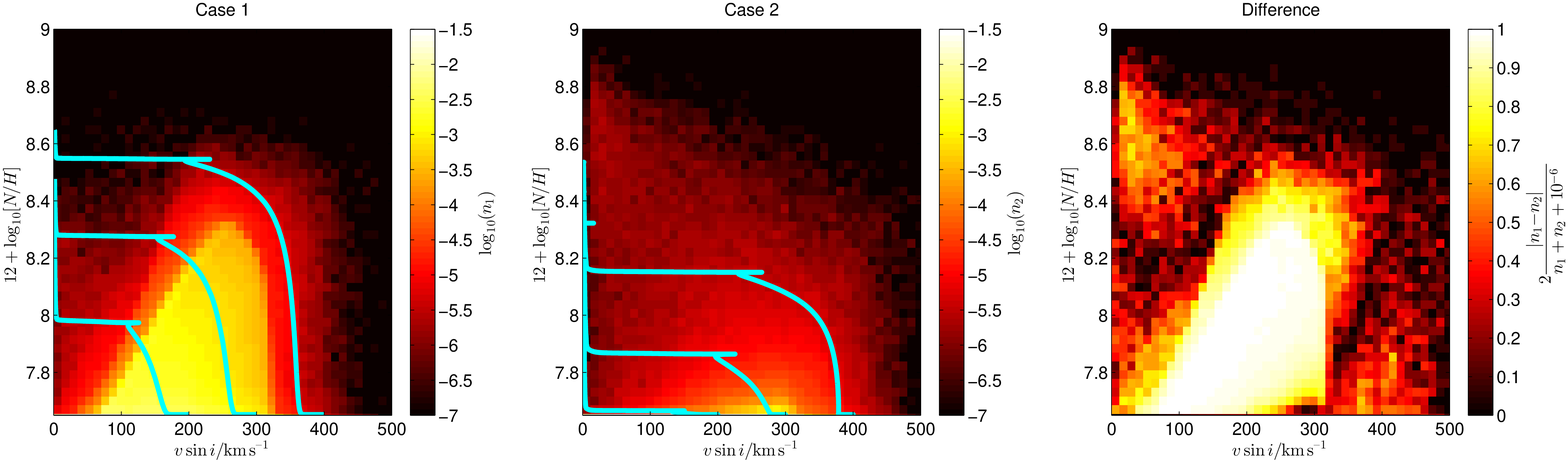}
\end{center}
\caption{Hunter diagrams for a population of stars with continuous star formation. In each case we have plotted the evolution of $10$\ms\ stars with initial surface rotation velocities $200\,{\rm km\,s^{-1}}$, $300\,{\rm km\,s^{-1}}$ and $400\,{\rm km\,s^{-1}}$. Despite showing good agreement at low surface rotation rates, case~1 has many more fast-rotating highly enriched stars. However, this difference can often be accounted for by recalibration of the mixing coefficients and is difficult to observe owing to the rarity of rapidly rotating high-mass stars which occupy this region of the plot.}
\label{huntcont}
\end{figure*}

If we consider the case dependence of the Hunter diagrams for a population of stars with a continuous star formation history we find much the same thing. Fig.~\ref{huntcont} shows that, although both cases follow a trend of more frequent enrichment in stars with higher rotation rates, case~1 produces far more enriched stars than case~2. This is not surprising because we found far less mixing in case-2 stars at lower masses than in case~1 and it is these stars that dominate the population. We could have instead chosen to calibrate case~2 so that there were more mixing in low-mass stars but this would inevitably lead to a worse match in the populations elsewhere, perhaps in the enrichment of rapidly rotating very massive stars ($M>40$\ms). We discuss this further in section \ref{calibration}. Also important to consider is the effect of metallicity. In Paper~1 we found a reversal in the trend of less mixing in case-2, low-mass stars ($M<20$\ms). Increasing the mixing here to bring the two populations in line would make the low-metallicity agreement far worse. We discuss this in section \ref{metal}.

\subsection{Effective surface gravity and enrichment}
\label{grav}

\begin{figure}
\begin{center}
\includegraphics[width=0.48\textwidth]{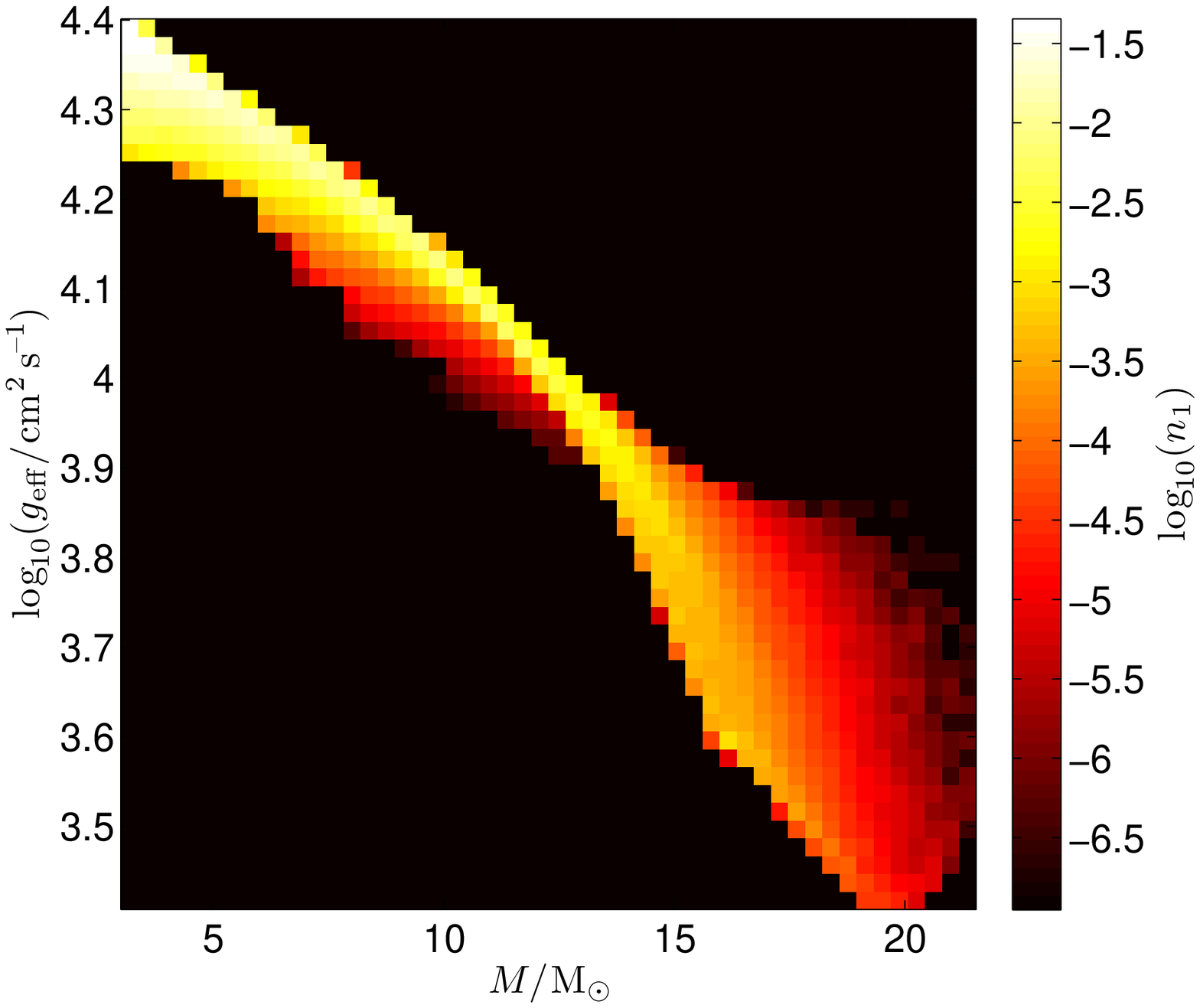}
\end{center}
\caption{Simulated single-aged stellar population at $10^7{\rm \,yr}$ in case~1. The plot shows the strong correlation between mass and surface gravity. This relation only holds when the population has a single age. Because of rotation, the relation is degenerate and a measurement of the surface gravity corresponds to stellar masses with a range of up to $7$\ms.}
\label{gravfig}
\end{figure}

\begin{figure*}
\begin{center}
\hspace*{-2.2cm}\includegraphics[width=1.2\textwidth]{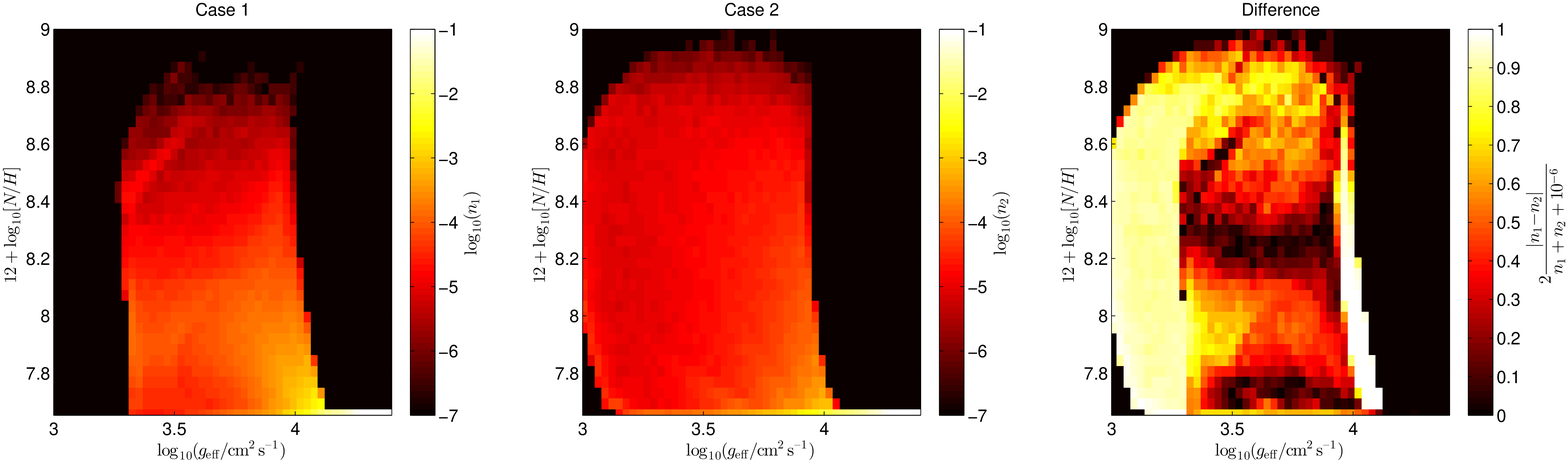}
\hspace*{-2.2cm}\includegraphics[width=1.2\textwidth]{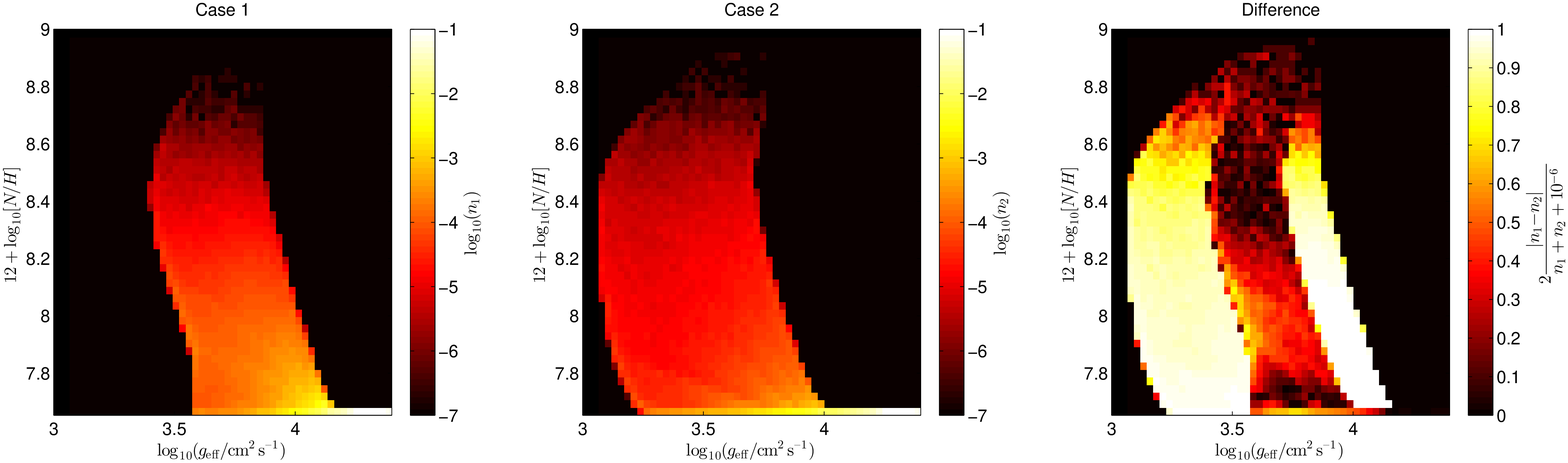}
\hspace*{-2.2cm}\includegraphics[width=1.2\textwidth]{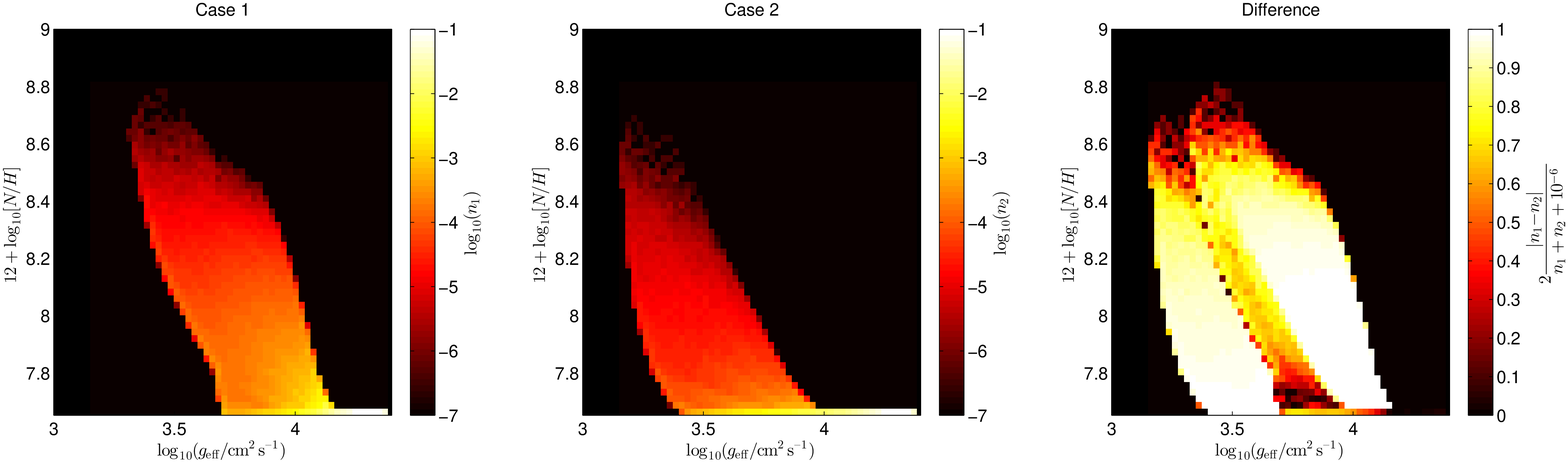}
\hspace*{-2.2cm}\includegraphics[width=1.2\textwidth]{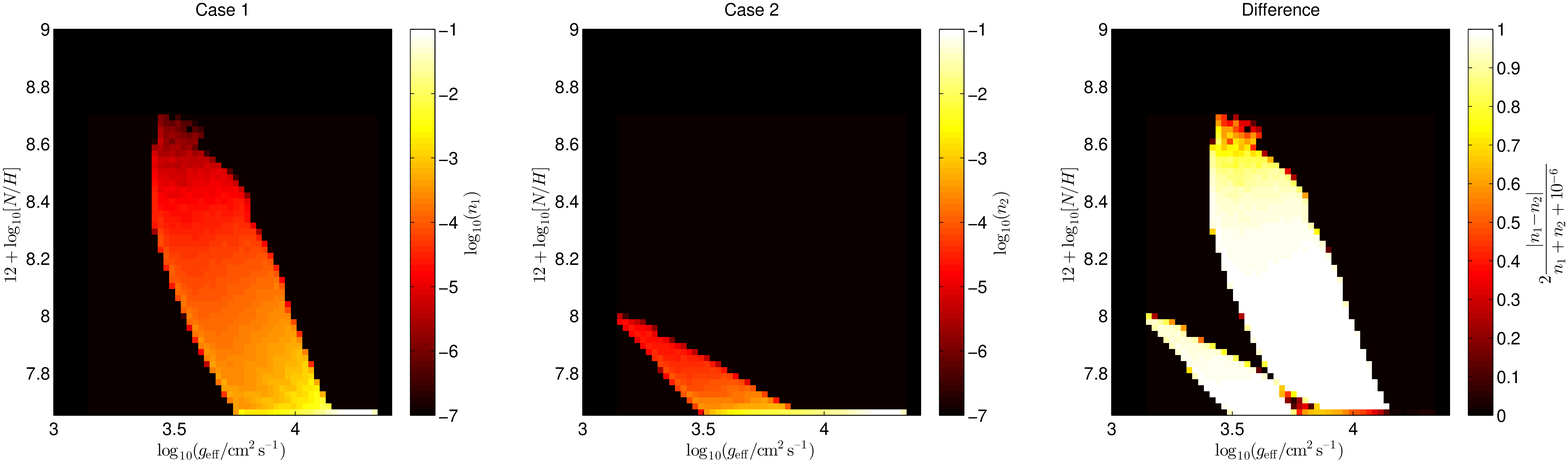}
\end{center}
\caption{Distribution of the surface nitrogen enrichment against effective surface gravity in single-aged stellar populations. From top to bottom, the four rows of figures correspond to $5\times 10^6 {\rm yr}$, $10^7 {\rm yr}$, $2\times 10^7 {\rm yr}$ and $5\times 10^7 {\rm yr}$. At early times case~2 gives a larger spread of effective surface gravities and higher enrichment of the fastest rotators. At later times the maximum enrichment is similar but case~2 predicts overall lower surface gravities than case~1. Finally at late times when only stars with mass smaller than $8.2$\ms$\ $ remain, case~2 predicts far less mixing than case~1 as well as much lower surface gravity for its fastest rotators. }
\label{gn}
\end{figure*}

As we suggested in Paper~1, the difference between the two cases can be seen most clearly by considering different masses of stars. In our discussion of the Hunter diagram we have appealed to the single-aged population of stars to differentiate between stellar masses as the population ages. Unfortunately, determining the mass of rotating stars self-consistently is difficult because of the degeneracy that arises owing to rotation. Fig.~\ref{gravfig} shows the typical relationship between mass and effective surface gravity in a simulated population. There is a strong correlation between the two but rotation causes degeneracy so estimates of the mass from effective gravity alone could be wrong by up to $7$\ms$\,$in this case. The correlation does not persist in the case of continuous star formation. Use of the effective surface gravity is also advantageous because it can be directly determined spectroscopically. However, caution is necessary for rapid rotators because the effective gravity is not uniform across the stellar surface \citep{VonZeipel24}. The Hunter diagram suffers from the problem that, even for simple stellar populations like this one, stars exist in all regions of the diagram and the population has few clear boundaries. If we look at the variation of effective surface gravity with nitrogen enrichment the difference between the models becomes very clear (Fig.~\ref{gn}). There are sharp curves that bound the upper and lower effective surface gravities of the population. The lower bound occurs because stars evolve rapidly into giants with much lower surface gravity after this limit. The upper bound occurs because younger stars with higher surface gravities haven't evolved to the point where their surface nitrogen is enriched. There are features which distinguish the two populations at each age. For young populations ($5\times 10^6{\rm yr}$) case~2 has a higher upper bound for nitrogen enrichment and there is a much broader range of surface gravities than in case~1. For older populations ($10^7{\rm yr}$ and $2\times 10^7{\rm yr}$) case~2 predicts generally lower values for the surface gravity. Finally for old populations ($5\times 10^7{\rm yr}$) the difference becomes very stark. The amount of mixing in case~2 drops off dramatically compared to case~1 while we still predict much lower values for the surface gravity in rapid rotators.

\begin{figure*}
\begin{center}
\hspace*{-2.2cm}\includegraphics[width=1.2\textwidth]{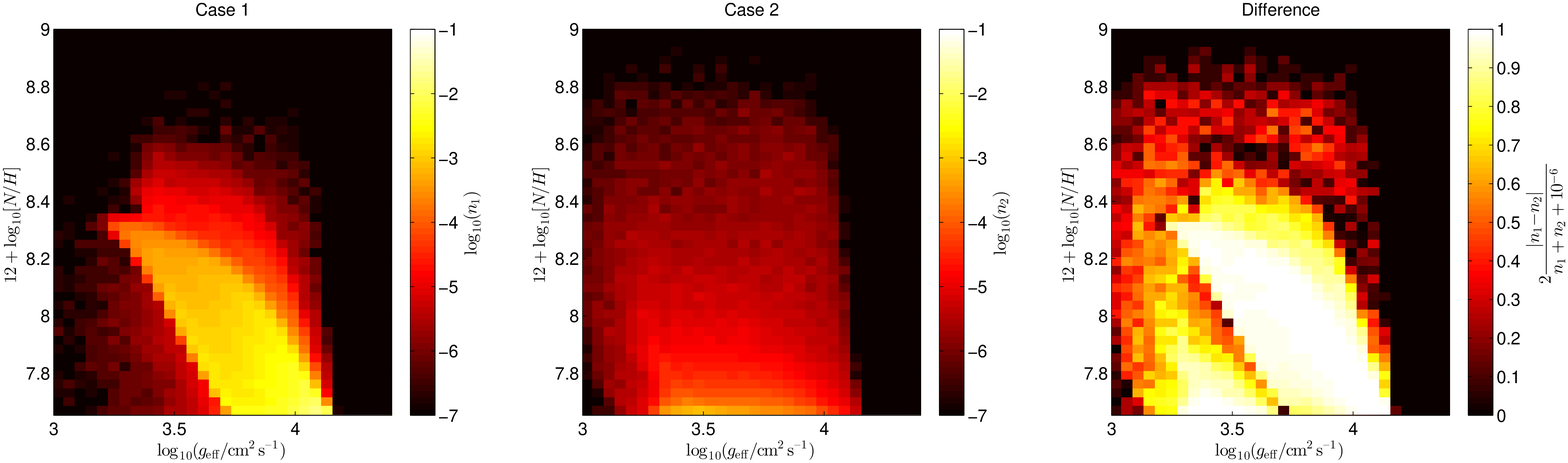}
\end{center}
\caption{Surface nitrogen enrichment against effective surface gravity of simulated populations of stars with continuous star formation. The distributions are both confined to a narrow band and have similar ranges for enrichment, though slightly higher in case~2. However, case~1 produces many more enriched stars than case~2 when they are for the large part confined to a narrow band. There are some edge-of-grid effects that arise because the initial rotation function is mass independent and so produces many more low-mass stars close to their critical limit.}
\label{contgn}
\end{figure*}

When we consider a population of stars with continuous star formation history, the difference in the populations is still clear. Interestingly, unlike the Hunter diagram, this visualisation actually highlights the similarities as well as the differences between the two cases. Fig.~\ref{contgn} shows that stars in both models are confined to a similar band of effective gravities and their range of surface abundances are very similar. The main difference between the two cases, apart from the increased frequency of enriched stars in case~1 which we saw in section \ref{huntsec}, is the confinement of the enriched case-1 stars to a distinct band. This contrasts to case~2 for which the stars are spread much more evenly across their range of enrichment.

\subsection{Recalibration}
\label{calibration}

\begin{figure*}
\begin{center}
\hspace*{-2.2cm}\includegraphics[width=1.2\textwidth]{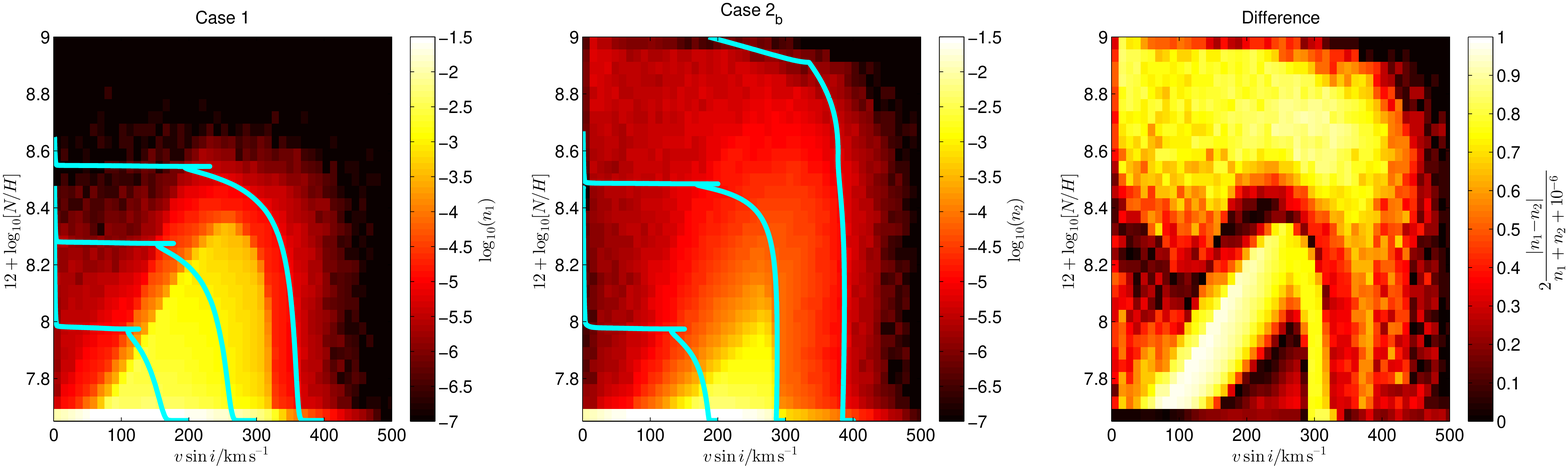}
\caption{Hunter diagrams for a population of stars undergoing continuous star formation with case~2 calibrated to give the same TAMS nitrogen enhancement as case~1 for a star of mass $10$\ms\ initially spinning with $v=200\,{\rm km\,s^{-1}}$ (case~2$_{\rm b}$). In each case we have plotted the evolution of $10$\ms\ stars with initial surface rotation velocities $200\,{\rm km\,s^{-1}}$, $300\,{\rm km\,s^{-1}}$ and $400\,{\rm km\,s^{-1}}$. There are now more moderately enriched stars in case~2$_{\rm b}$ but still far fewer than in case~1 and the upper limit for enrichment in case~2$_{\rm b}$ is now far greater than case~1.}
\label{huntcontb}
\end{center}
\end{figure*}

We have thus far described the differences that arise between the two test cases under a specific calibration of the mixing. However, within each case there is the flexibility to calibrate to some degree the amount of mixing that arises because of rotation. We chose in our initial calibration to match the TAMS nitrogen enrichment of $20$\ms$\ $ stars initially rotating at $v=300\,{\rm km\,s^{-1}}$. This is a reasonably good fit for stars of $M>15$\ms\ but for smaller masses the amount of mixing in case~2 drops off rapidly. Now suppose instead we had chosen to match the TAMS nitrogen enrichment of a star with initial mass $M=10$\ms$\,$ and $v=200\,{\rm km\,s^{-1}}$. We refer to this model as case 2$_{\rm b}$. This is more representative of the stars observed in the VLT-FLAMES survey \citep{Dufton06} and so should produce mixing in line with the bulk of the population. We discuss the VLT-FLAMES survey data in relation to our simulated populations in section \ref{flames}.  Fig.~\ref{huntcontb} shows the Hunter diagram for the new sample. We see that the agreement is better in the Hunter diagram but case~2$_{\rm b}$ still can't produce the tightly confined bulk of enriched stars seen in case~1. Also, the maximum enrichment observed in case~2$_{\rm b}$ is now far greater than in case~1.

\begin{figure*}
\begin{center}
\hspace*{-2.2cm}\includegraphics[width=1.2\textwidth]{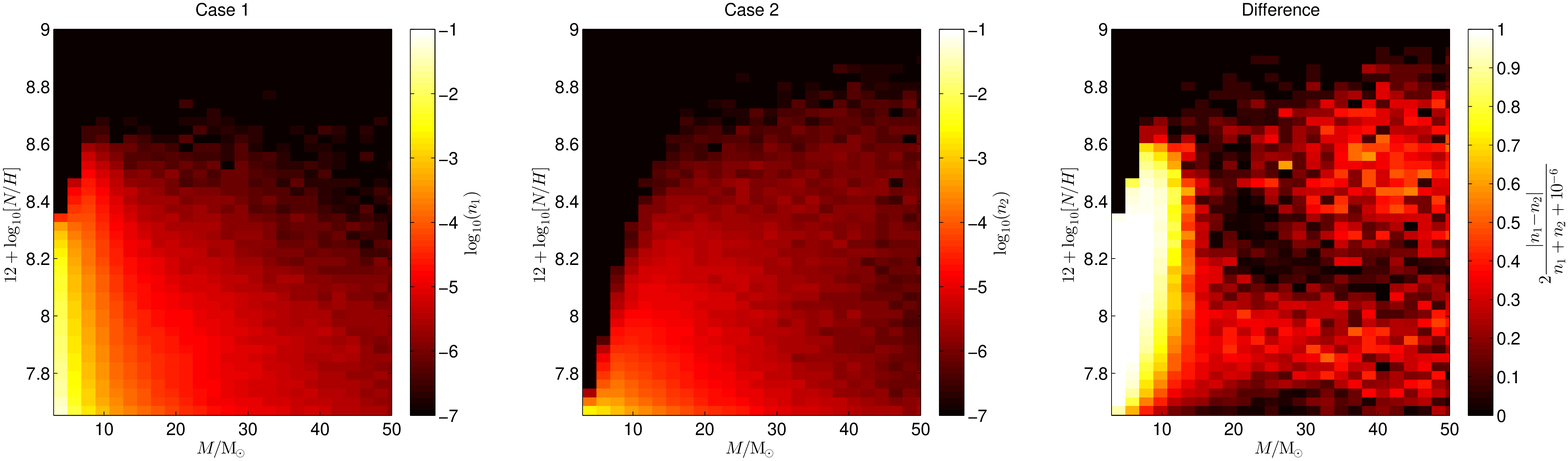}
\hspace*{-2.2cm}\includegraphics[width=1.2\textwidth]{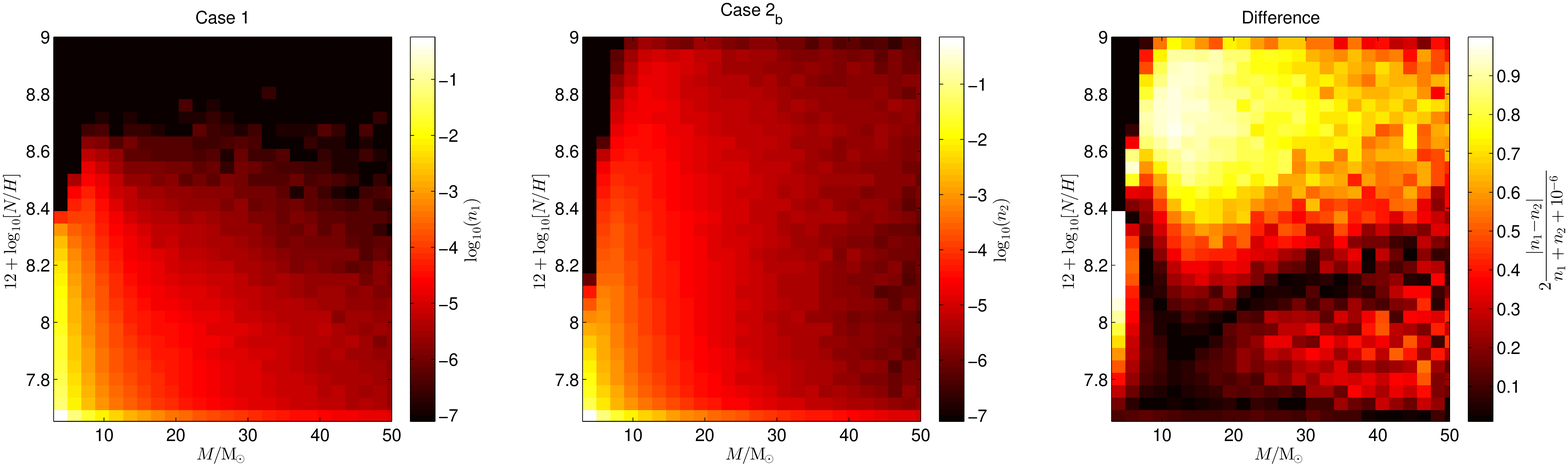}
\end{center}
\caption{Distribution of stars in case~1 and case~2 for a population with continuous star formation history. The top row shows the models with the calibration of case~2 to give the same TAMS nitrogen enhancement as case~1 for a star of $20$\ms\ initially spinning with $v=300\,{\rm km\,s^{-1}}$. The second row shows the same but with case~2 calibrated to give the same TAMS nitrogen enrichment as as star with initial mass $M=10$\ms$\ $ and surface rotation $v=200\,{\rm km\,s^{-1}}$ (case~2$_{\rm b}$). The two cases agree for masses greater than  $15$\ms$\ $ for the original calibration. The agreement continues to lower masses for the second calibration but now there are more highly enriched stars in case~2$_{\rm b}$ in the mass range $5$\ms$<M<20$\ms.}
\label{mn}
\end{figure*}

Although we can't directly measure the mass of stars, it is instructive to examine where the main differences in our sample arise. Fig.~\ref{mn} shows the distribution of nitrogen enrichment by mass in case~1, case~2 and case~2$_{\rm b}$. In the first instance, the agreement between the two models is reasonable for stars more massive than around $15$\ms$\ $ but the mixing in case~2 drops off rapidly for lower masses as we observed in section \ref{grav}. For case~2$_{\rm b}$ the agreement holds to much lower masses except now there are far more highly enriched stars with mass $5$\ms$<M<20$\ms\ when compared to case~1.

\subsection{Effects of metallicity}
\label{metal}

\begin{figure*}
\begin{center}
\hspace*{-2.2cm}\includegraphics[width=1.2\textwidth]{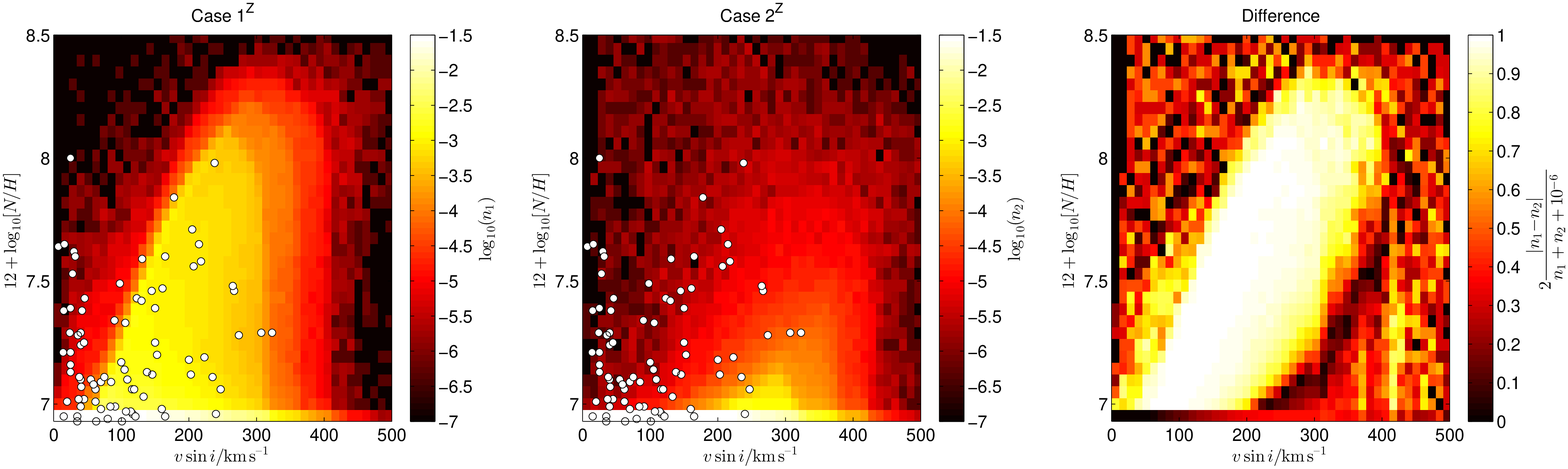}
\end{center}
\caption{Hunter diagrams for populations of stars undergoing continuous star formation at LMC metallicity. The two populations are qualitatively similar to the populations at solar metallicity. We have also plotted the LMC stars observed by \citet{Hunter09}. Case~1$^{\rm Z}$ predicts a very confined distribution of enriched stars, whereas case~2$^{\rm Z}$ predicts a much wider spread of enrichment. The stars at the left-hand edge of the diagram can't be explained by rotational mixing alone.}
\label{huntcontz}
\end{figure*}

In Paper~1 we concluded that there was significant variation in the differences between the two cases at different metallicities. We simulated a grid of models at LMC metallicity, composition and initial velocity distribution, as given by \citet{Brott11b}. This is not as low as the low-metallicity case described in Paper~1 but it does allow us to compare our results with the data from the LMC observations in the VLT-FLAMES survey of massive stars. We simulated a population in case~1$^{\rm Z}$ and case~2$^{\rm Z}$ at this composition with continuous star formation history. The Hunter diagram for this population is shown in Fig.~\ref{huntcontz}.

We see that the qualitative distribution of stars in the simulated population is similar to the solar metallicity populations. Case~1$^{\rm Z}$ produces a much more well-defined band of enriched stars whereas case~2$^{\rm Z}$ produces fewer enriched stars that have a much greater spread in abundance. If we consider the mass-dependence of the rotational mixing we find a similar decline in the amount of mixing in case~2$^{\rm Z}$ compared to case~1$^{\rm Z}$ for stars less massive than $20$\ms. As in Paper~1, we find that the amount of mixing in stars above this mass is higher in case~2$^{\rm Z}$ than in case~1$^{\rm Z}$. In fact, the mixing in case~1$^{\rm Z}$ decreases slightly for higher-mass stars. This means that, as in section \ref{calibration}, an increase in the mixing in case~2$^{\rm Z}$ is unlikely to produce a better correlation between the two cases. However, owing to the IMF, there are many more stars less massive than $20$\ms\ in the population, case~1 produces many more enriched stars than case~2$^{\rm Z}$.

In Fig.~\ref{huntcontz} we have also plotted those LMC stars for which the nitrogen abundances have been determined by \citet{Hunter09}. As remarked by \citet{Hunter09}, there are many highly enriched, slowly rotating stars that are not explained by either model of rotational mixing. However, for the remainder of observed stars we see a trend of increasing enrichment for higher rotation rates. On initial inspection, case~1$^{\rm Z}$ fits the VLT-FLAMES data much more closely than case~2$^{\rm Z}$. However, selection effects are important and we consider these in section \ref{flames}.

\subsection{Selection effects in the VLT-FLAMES survey}
\label{flames}

\begin{figure*}
\begin{center}
\hspace*{-2.2cm}\includegraphics[width=1.2\textwidth]{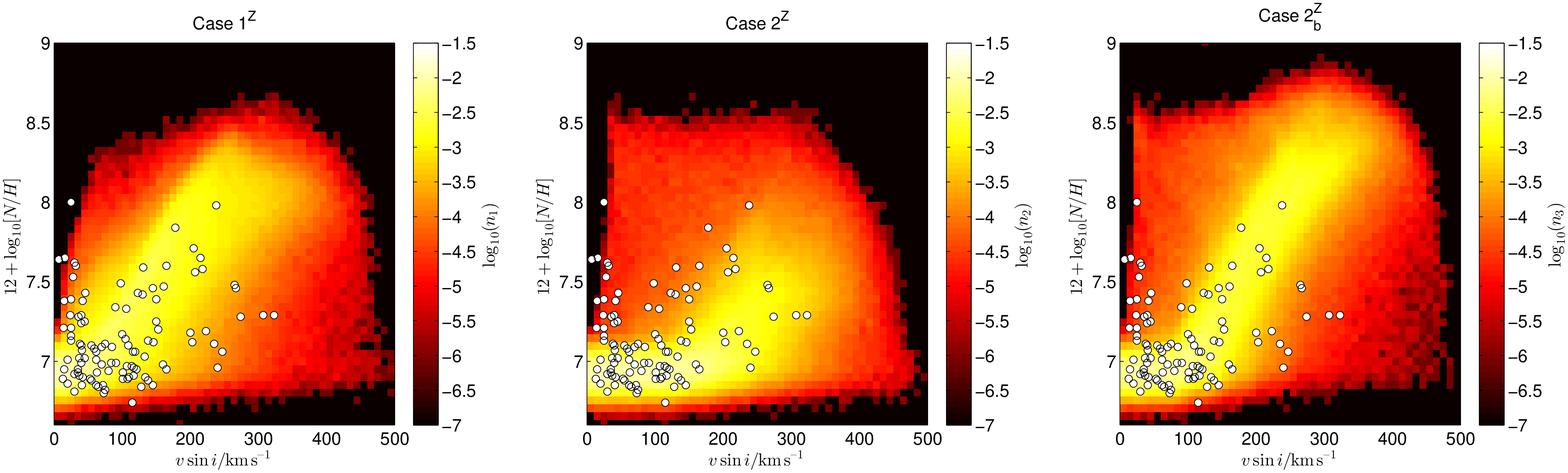}
\end{center}
\caption{Hunter diagrams for populations of stars undergoing continuous star formation at LMC metallicity for a number of different models with selection effects applied as described in section \ref{flames}. Populations have been simulated for cases~1$^{\rm Z}$, 2$^{\rm Z}$ and~2$_{\rm b}^{\rm Z}$. The populations are qualitatively similar to the populations at solar metallicity. We have also plotted the LMC stars observed by \citet{Hunter09}. The slowly-rotating, highly-enriched stars at the left-hand edge of the diagram cannot be explained by rotational mixing alone.}
\label{huntcontobsz}
\end{figure*}
\begin{figure}
\begin{center}
\includegraphics[width=0.48\textwidth]{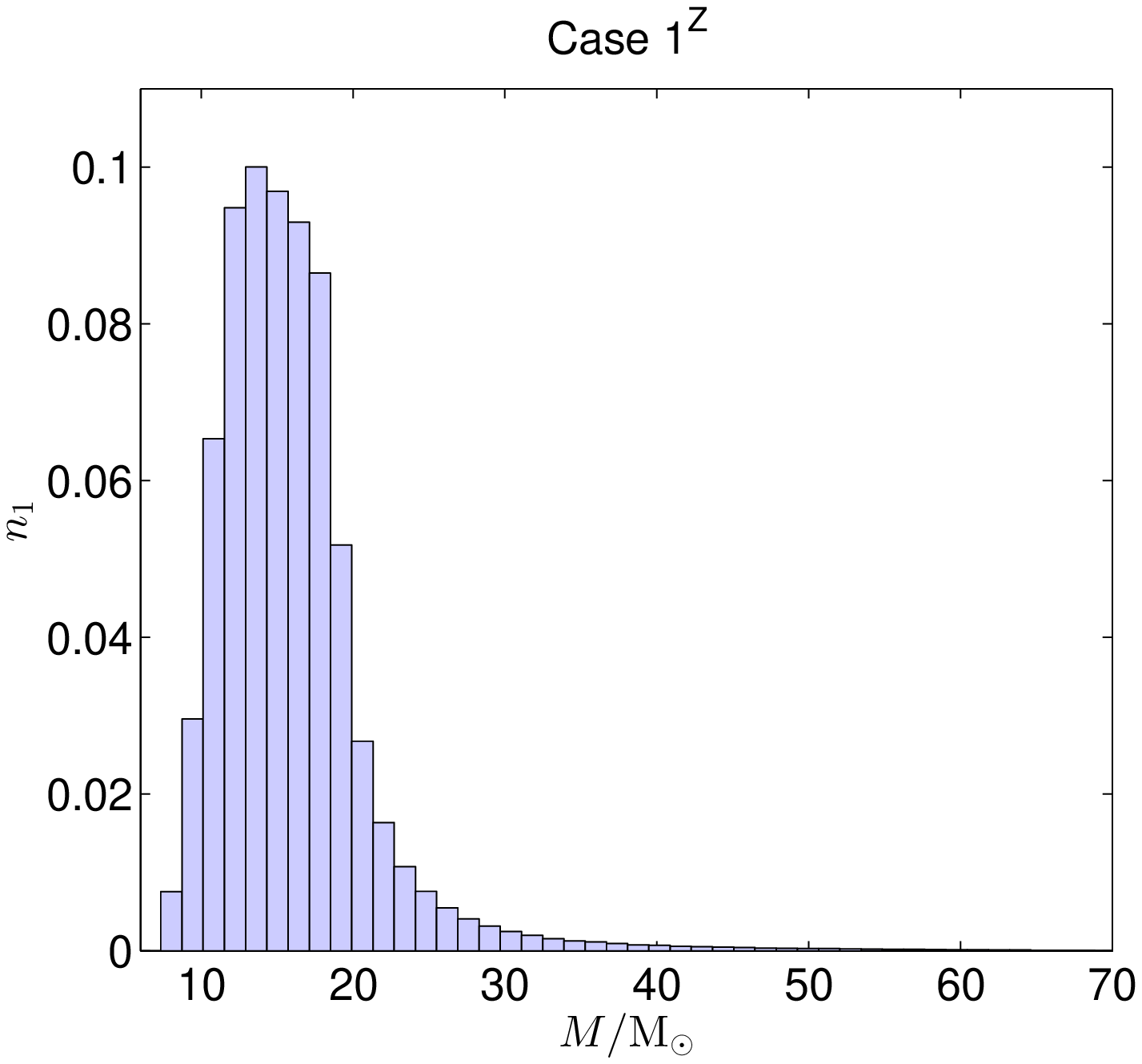}
\end{center}
\caption{Distribution of masses in a simulated population of case-1$^{\rm Z}$ stars at LMC metallicity with the inclusion of selection effects. The distribution is very similar in each case. We see that the distribution peaks strongly around $12$\ms.}
\label{masscontobsz}
\end{figure}

The most common data set used to test rotational mixing in massive stars is the VLT-FLAMES survey of massive stars \citep{Evans05,Evans06,Dufton06} owing to the number of stars sampled and the detailed determination of surface composition. We repeated our population synthesis as in section \ref{metal} for a continuous population of LMC-metallicity stars but we have included the selection effects which affect the stars in the VLT-FLAMES survey so that we may compare the distributions more directly with those found by \citet{Hunter09}. The selection criteria we used are that of the cluster N11 in the LMC. For a detailed description see \citet{Brott11}. Stars are excluded if their visual magnitude is greater than $15.34$, if they are hotter than $35{,}000$\,K, if their surface gravity is less than $10^{3.2}{\,\rm cm\,s^{-2}}$ or they are rotating faster than $90\%$ of their critical rotation rate. In addition, a random error is applied to $\log_{10}[N/H]$ selected randomly from a Gaussian with standard deviation $\sigma = 0.2$.

The simulated population produced after we apply the selection effects is shown in Fig. \ref{huntcontobsz} along with the LMC stars data of \citet{Hunter09}. We show the population in cases~1$^{\rm Z}$, 2$^{\rm Z}$ and~2$_{\rm b}^{\rm Z}$ (the LMC analogue of case~2$_{\rm b}$ with the calibration described in section \ref{calibration}). We see that, contrary to our discussion in section \ref{metal}, the differences between the various models are now far less apparent. Compared with the other two cases, case~2$^{\rm Z}$ predicts many more less-enriched fast rotators than are observed and the amount of mixing is insufficient to match the observed band of enriched stars. Compared with section \ref{metal}, cases~1$^Z$ and~2$^Z_b$ now both show a similarly good fit to the data. In both cases the band of predicted enriched stars is matched well by the observations. However, we note that we would expect to see a number of stars with $12+\log_{10}[N/H]>8$. The number of predicted stars in this range is greater for case~2$_{\rm b}^{\rm Z}$ than case~1$^Z$. If we were to reduce the amount of mixing we would get too little nitrogen enrichment in stars with $v<200{\rm \,km\,s^{-1}}$. This effect may be a result of the difficulty of measuring nitrogen abundances in this region but is otherwise difficult to resolve. A further increase in the mixing would exacerbate the problem that the upper bound to enrichment at rapid rotation is too high. A decrease in the mixing would mean that the band of enriched stars in the simulated population is less likely to produce a good fit to the model and would leave observed slowly-rotating, moderately-enriched stars that cannot be explained through the theoretical models.

Unfortunately, the mass dependence of the models is not well reflected in the VLT-FLAMES populations. With the selection effects described, the masses of the LMC-metallicity sample are confined between  $10$\ms\ and  $20$\ms\ as shown in Fig. \ref{masscontobsz}. We find a similarly narrow range when we consider Galactic stars under similar selection effects. Therefore, any simulated population where we include the selection effects of VLT-FLAMES captures the rotational-dependence of the model but for only a small fraction of the mass-dependence which is where we have found the biggest differences between our two cases.

\section{Conclusions}

Rotation has many effects on stellar evolution. Some of these, such as that on the surface temperature, are because of rotation but only vary significantly between different models towards the end of the main sequence. Others, such as that on the surface rotation velocity, may evolve differently throughout the main sequence according to different models but do not produce significant changes in the distribution of stars in a simulated population. Therefore these properties alone are largely unhelpful to distinguish between the different implementations of rotation in stellar models. 

It has been observed that the surface abundances of several chemical elements change significantly because of rotation. The degree to which this happens in the theoretical models strongly depends on which particular model for stellar rotation is used and which constraints are used to calibrate them. Recently, the Hunter diagram has been the favoured diagnostic tool for analysing stellar rotation because it shows a clear connection between the surface rotation of a star and its surface enrichment. Our model based on that of \citet{Talon97}, case~1, shows a similar order of magnitude enrichment for all masses. On the other hand our model based on that of \citet{Heger00}, case~2, shows a steep decline in the amount of enrichment around $15$\ms. We can account for this to a degree by adjusting the calibration of the models but we see that increasing the mixing in case~2, so that low-mass stars show similar enrichment to those of case~1, then leads to a much higher maximum enrichment for a case-2 population than a case-1 population. This suggests that studies should focus on stars either side of this mass limit.

Because the two models are very different for different mass ranges, the effective gravity is a sensible tool to investigate the mass-dependence of the mixing strength. It is very difficult to self-consistently infer the mass of a star from a luminosity--temperature--rotation relation because any such relation depends on the model used for rotation. The effective gravity however can be derived directly from spectra and, despite some degeneracy, there is a strong relation between it and the stellar mass. This can be usefully applied to the study of rotational mixing. We have shown that there are very clear differences between the synthetic populations produced by each model when nitrogen enrichment is plotted against surface gravity. In the case of a population in which all the stars have the same age, they are confined to a very specific region. Stars with surface gravity below a certain limit evolve into giants and beyond on a short timescale compared to their main-sequence lifetime. Stars with surface gravity above another limit are less massive and have not had long enough to become enriched. This effect persists even in the case of a population of stars with continuous star formation. Case~1 predicts that nitrogen is enriched over a narrower range of surface gravities than case~2. We see that case~1 produces many more moderately enriched stars than case~2 but the maximum enrichment in case~1 is lower than in case~2.

We have also shown that similar trends appear at different metallicities. We simulated populations in both cases with continuous star formation history at LMC metallicity. The qualitative distribution of stars in each case was similar to that at Galactic metallicity. Case~1$^{\rm Z}$ still produced a confined band of enriched stars in the Hunter diagram, whereas case~2$^{\rm Z}$ produced a much greater spread. Case~1$^{\rm Z}$ also produced many more highly-enriched stars than case~2$^{\rm Z}$ although the maximum enrichment in case~2$^{\rm Z}$ was higher than in case~1$^{\rm Z}$. Similarly to solar metallicity, this is because of the mass dependence of the mixing in each case. The decline in the amount of mixing in case~2$^{\rm Z}$ begins at higher masses at low metallicity ($20$\ms\ at LMC rather than $15$\ms$\,$ at solar metallicity) and the mixing in stars above this mass is relatively constant for stars in case~2$^{\rm Z}$, whereas case-1$^{\rm Z}$ stars show slightly less enrichment as the mass of stars increases. This may not be indicative of the strength of the rotational mixing but rather that, because the main-sequence lifetime of the stars decreases with increasing mass, there is less time to transport nitrogen from the core to the surface. 

When we compare the simulated populations to the LMC-metallicity stars in the VLT-FLAMES survey we find that both cases~1$^{\rm Z}$ and~2$_{\rm b}^{\rm Z}$, which uses the second calibration for case~2 described in section \ref{calibration}, give a reasonable fit to the observed data and it is difficult to determine which fits the data more closely. Unfortunately the range of initial masses that remain in the simulated samples after we apply the selection effects is extremely narrow, between $10$\ms\ and $20$\ms\ in the LMC populations. This means that it is difficult to observe the large difference in the mass-dependence of the models. Hence, a close fit between the Hunter diagrams for a simulated population and the data of the VLT-FLAMES survey is a useful test of models for rotational mixing but cannot establish the validity of a model by itself. As we have seen, models for rotational mixing with very different mass-dependencies can reproduce similarly good fits to the current data. Whilst, in the future, determinations of the nitrogen abundance in a wider mass-range of stars may help solve this problem, it is likely that examining the enrichment and depletion of other elements will be necessary. For example, \citet{Brott11b} look at the effect of rotation on the surface Boron abundance. Different initial rotation velocity distributions may also have a significant effect on the simulated populations. For example an initial distribution which is a function of $\Omega/\Omega_{\rm crit}$ would prevent the over--abundance of rapid rotators at low masses.

Despite sharing many similar features, it is unreasonable to expect that two different models for stellar rotation can produce identical qualitative results for an extended range of masses, rotation rates and metallicities. We have shown that, whilst the two models agree for stars more massive than $15$\ms, there is much less agreement for less massive stars. Furthermore, we have only thus far made a comparison of two particular models for stellar rotation chosen from the many available. In particular we haven't yet included models based around the Taylor-Spruit dynamo \citep{Spruit02} such as that investigated by \citet{Brott11b}. Whilst this is a similar model to that of \citet{Heger00} it produces very different results owing to the inclusion of magnetic fields. This extension is supported by our consideration of different metallicity regimes. Whilst the better fit of case~1 suggests that models in which meridional circulation is treated advectively and diffusion comes solely from hydrodynamical instabilities are more realistic, the results of \cite{Brott11} also produce a reasonable fit to observed LMC stars. To distinguish between these models, and others, it is necessary to continue with this analysis and extend it to different masses and metallicities as more data becomes available.

\section{Acknowledgements}
ATP thanks the STFC for his studentship and CAT thanks Churchill college for his fellowship.

\bibliographystyle{mn2e}
\bibliography{paper8}

\end{document}